\documentclass[conference,letterpaper]{IEEEtran}
\IEEEoverridecommandlockouts

\usepackage[utf8]{inputenc} 
\usepackage[T1]{fontenc}
\usepackage{url}
\usepackage{ifthen}
\usepackage{multicol}
\usepackage{flushend}
\usepackage{balance}
\usepackage[cmex10]{amsmath} % Use the [cmex10] option to ensure complicance
\newtheorem{theorem}{Theorem}

\newtheorem{lemma}{Lemma}

\newtheorem{corollary}{Corollary}

\newcommand{\indep}{\perp\!\!\!\perp}

\usepackage{tikz}
\usepackage{caption}

\usepackage[style=ieee]{biblatex}
\usepackage{amssymb}
\usepackage{physics}
\usepackage{algorithm}
\usepackage[noend]{algpseudocode}
\usepackage{subfiles}
\title{Broadcast Channel Synthesis from Shared Randomness \thanks{The work of M. Managoli and V. Prabhakaran was supported by DAE under project no. RTI4001. V. Prabhakaran was additionally supported by SERB through project MTR/2020/000308.}}

\author{%
	\IEEEauthorblockN{Malhar A. Managoli\IEEEauthorrefmark{1} and Vinod M. Prabhakaran\IEEEauthorrefmark{2}}
	\IEEEauthorblockA{School of Technology and Computer Science\\
		Tata Institute of Fundamental Research\\
		Mumbai, India\\
		Email: \IEEEauthorrefmark{1}malhar.managoli@tifr.res.in,  \IEEEauthorrefmark{2}vinodmp@tifr.res.in}
}
\begin{document}
\maketitle
\begin{abstract}
	We study the problem of synthesising a two-user broadcast channel using a common message, where each output terminal shares an independent source of randomness with the input terminal. This generalises two problems studied in the literature (Cuff, \emph{IEEE Trans. Inform. Theory}, 2013; Kurri et.al., \emph{IEEE Trans. Inform. Theory}, 2021). We give an inner bound on the tradeoff region between the rates of communication and shared randomness, and a lower bound on the minimum communication rate. Although the bounds presented here are not tight in general, they are tight for some special cases, including the aforementioned problems.
\end{abstract}
\section{Introduction}

%%intro
\subsection{Motivation and Related Works}
In the {\em channel synthesis} problem for a channel $q_{Y|X}$, an input terminal receives independent and identically distributed (i.i.d.) copies of a random variable $X$ distributed according to $q_X$ and an output terminal is required to output copies of the random variable $Y$ such that the inputs and outputs are approximately i.i.d. according to $q_Xq_{Y|X}$. Variants with multiple input and/or output terminals have also been studied; this problem is also known as the {\em coordination} problem. Three different notions of how the actual distribution approximates the desired one have been considered in the literature - weak coordination~\cite{cuff_coordination, letreust2015, cervia2016, mylonakis2019empirical}, strong coordination~\cite{wyner1975common, bennett, harsha, cuff, bennett2014quantum, yassaee15}, and exact coordination~\cite{kumar, vellambi, vellambi2018}. In this paper, we focus on strong coordination, which requires that the total variation distance between the synthesized distribution and the desired distribution vanishes as the number of copies (block length) goes to $\infty$.

Bennett et al.~\cite{bennett} studied the point-to-point channel synthesis problem when the terminal share unlimited common randomness. Cuff~\cite{cuff} considered the benefit of limited common randomness, characterising the trade-off between communication and common randomness. Synthesis of a broadcast channel using limited randomness common to all terminals was also studied in~\cite{cuff}. The optimal trade-off between the rate of a common message from the input terminal to all output terminals and the rate of common randomness shared by all terminals was obtained there. 

\begin{figure}[H]
	\centering
	\begin{tikzpicture}[scale=2]
		\draw (0,-.55) node{$X^n$};
		\draw[->] (0,-.7) -- (0,-1.2);
		\draw (-.5,-1.2) rectangle (.5,-1.7);
		\draw (0,-1.45) node{$P_1$};
		\draw (0,-1.7) -- (0,-2.7);
		\draw[->] (0,-2.7) -- (-.75,-2.7);
		\draw[->] (0,-2.7) -- (.75,-2.7);
		\draw (0,-2.9) node{$M$};
		\draw (-.75,-2.95) rectangle (-1.75,-2.45);
		\draw (.75,-2.95) rectangle (1.75,-2.45);
		\draw (-1.25,-2.7) node{$P_2$};
		\draw (1.25,-2.7) node{$P_3$};
		\draw[->] (-1.25,-2.95) -- (-1.25,-3.45);
		\draw[->] (1.25,-2.95) -- (1.25,-3.45);
		\draw (-1.25,-3.6) node{$Y^n$};
		\draw (1.25,-3.6) node{$Z^n$};
		\draw[dashed,<->] (-.5,-1.45) to[out=200,in=90] (-1.25,-2.45);
		\draw[dashed,<->] (.5,-1.45) to[out=340,in=90] (1.25,-2.45);
		\draw (-.875,-1.95) node{$\Theta_1$};
		\draw (.875,-1.95) node{$\Theta_2$};
	\end{tikzpicture}
	\caption{Synthesis of a broadcast channel. $P_1$ observes $X^n\sim q_X$ i.i.d.. $P_2$ and $P_3$ must output $Y^n$ and $Z^n$ respectively, such that the joint distribution is close to $q_X\,q_{Y,Z|X}$ i.i.d.. $P_1$ can send an $nR$ bits long common message to $P_2$ and $P_3$. $P_1$ and $P_2$ have access to $nR_1$ uniformly distributed random bits $\Theta_1$. Similarly, $P_1$ and $P_3$ have access to $nR_2$ uniformly distributed random bits $\Theta_2$. Here $X^n,\Theta_1,\Theta_2$ are independent.}
	\label{fig:prob}
\end{figure}
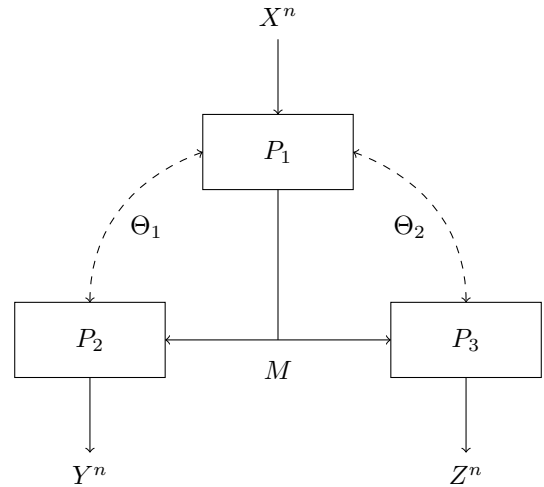

Motivated by the fact that even if the terminals do not have access to a \textit{common} source of randomness, subsets of them might share randomness, we consider the benefit of such forms of {\em shared} randomness in this paper. Specifically, we consider the synthesis of a broadcast channel with two output terminals when the input terminal shares independent randomness with each of the output terminals (see Figure~\ref{fig:prob}). Coordination~\cite{gowtham} and multiple access channel synthesis~\cite{gowtham22} have been studied under this model.

We give an inner bound on the achievable region of rates of communication and shared randomness; and a lower bound on the minimum communication rate, when unlimited shared randomness is available.
The bounds for the minimum communication rate leave a gap in general, but are tight for several special cases. Some of these special cases have been studied before and we recover these. We also have a new result for the special case $Y=Z$ (i.e., the decoders' outputs have to be identical).

\subsection{Notation}
\begin{itemize}
	\item $[m]$ refers to the set $\{1,2,\ldots,m\}$
	\item Random variables are denoted by capital letters ($X,Y,Z$ etc.). Their realisations are denoted by small letters ($x,y,z$ etc.). Supports of random variables are denoted by corresponding calligraphic letters ($\mathcal{X},\mathcal{Y},\mathcal{Z}$ etc.).
	\item Sequences of random variables like $(X_1,X_2,\ldots,X_n)$ are denoted by $X^n$.
	\item Random probability mass functions (p.m.f.) are denoted by capital letters ($P_X,Q_{XY}$, etc.). Non-random p.m.f.s are denoted by small letters ($p_X,q_{XY}$, etc.). When evaluating p.m.f.s, we use $p(x,y)$ as shorthand for $p_{XY}(x,y)$, etc.
	\item For $0\leq p\leq 1$, the binary entropy function $h(p)$ is
	\begin{align*}
		h(p)=p\log\frac{1}{p}+(1-p)\log\frac{1}{1-p}
	\end{align*}
\end{itemize}

\section{Problem Statement}

%% Problem Statement

%\begin{itemize}
%	\item Three agents $X$, $Y$, $Z$ (abuse of notation).
%	\item $X$ can send a common message $M$ to $Y$ and $Z$.
%	\item $X,Y$ have access to shared randomness $\Theta_1$.
%	\item $X,Z$ have access to shared randomness $\Theta_2$.
%	\item Required to simulate the broadcast channel $q_{YZ|X}$ with input distribution $q_X$.
%\end{itemize}
Let $\mathcal{X}$, $\mathcal{Y}$, $\mathcal{Z}$ be finite sets (alphabets) and let $P_1$, $P_2$, $P_3$ be three terminals. $P_1$ observes $n$ i.i.d. copies of a random variable $X$ over $\mathcal{X}$ according to a known distribution $q_X$ and can send a $nR$ bits long common message $M$ to $P_2$ and $P_3$. Further, $P_1$ and $P_2$ have $nR_1$ bits of shared randomness $\Theta_1\sim\mathrm{Unif}\{1,\ldots,2^{nR_1}\}$, and $P_1$ and $P_3$ have $nR_2$ bits of shared randomness $\Theta_2\sim\mathrm{Unif}\{1,\ldots,2^{nR_2}\}$. $X^n,\Theta_1,\Theta_2$ are independent. All terminals can privately randomise.

Together the terminals want to synthesise a broadcast channel $q_{YZ|X}$. $P_2$ has to produce $Y^n$ based on $M,\Theta_1$, and $P_3$ has to produce $Z^n$ based on $M,\Theta_2$, such that the resulting distribution $p_{X^nY^nZ^n}$ is close to $\prod_{i=1}^{n}q_X\,q_{YZ|X}$ in total variation.

The joint distribution of $X^n,Y^n,Z^n,M,\Theta_1,\Theta_2$ is of the form $q_{\Theta_1}q_{\Theta_2}q^n_X \, p_{M|X^n\Theta_1\Theta_2}\, p_{Y^n|M\Theta_1}\, p_{Z^n|M\Theta_2}$, where $q_{\Theta_1},q_{\Theta_2}$ are uniform p.m.f.s over $\{1,\ldots,2^{nR_1}\}$ and $\{1,\ldots,2^{nR_2}\}$ respectively. $p_{M|X^n\Theta_1\Theta_2}$ is the (privately randomised) encoding function, and $\ p_{Y^n|M\Theta_1}$ and $p_{Z^n|M\Theta_2}$ are the (privately randomised) decoding functions.

The triple of conditional distributions $p_{M|X^n\Theta_1\Theta_2},\ p_{Y^n|M\Theta_1}$, and $p_{Z^n|M\Theta_2}$ define an $(R,R_1,R_2,n)$-code. 
%These can be thought of as (randomised) encoding and decoding functions.
A tuple $(R,R_1,R_2)$ is said to be \emph{achievable} if and only if there exists a sequence of $(R,R_1,R_2,n)$-codes for which the following holds:
\begin{align}
	\lim_{n\rightarrow\infty}\norm{p_{X^nY^nZ^n}- \prod_{i=1}^{n}q_Xq_{YZ|X}}_{\mathrm{TV}}=0 \label{eq:2}
\end{align}
The \emph{rate region} $\mathcal{R}$ is the closure of the set of all achievable rate tuples.

The \emph{minimum communication rate} $R_{\mathrm{opt}}$ is the infimum of all rates $R$, such that $\exists R_1,R_2\ge0$ with $(R,R_1,R_2)\in\mathcal{R}$.

\section{Results}

%%results
\subsection{Inner Bound}

\begin{theorem}
	\label{th:ach}
	Define the rate region
	\begin{align*}
		\mathcal{S}={}&\begin{cases}
			(R,R_1,R_2)\in\mathbb{R}^3_{\ge0}:\ \exists\, p_{XYZUVW}\in\mathcal{D}\ \mathrm{s.t.}\\
			R\ge I(V;W|U)+I(U,V,W;X)\\
			2R\ge I(V;W|U)+I(U,V,W;X)+I(U;X,Y,Z)\\
			R+R_1\ge I(U,V;X,Y,Z)\\
			R+R_2\ge I(U,W;X,Y,Z)\\
			R+R_1+R_2\ge I(V;W|U)+I(U,V,W;X,Y,Z)\\
			2R+R_1+R_2\ge I(V;W|U)+I(U,V,W;X,Y,Z)\\
			\qquad\qquad\qquad\qquad\!\!\!+I(U;X,Y,Z)
		\end{cases}
	\end{align*}
	where
	\begin{align*}
		\mathcal{D}={}&\left\{
			p_{XYZUVW}={} q_X \:p_{UVW|X}\: p_{Y|UV}\: p_{Z|UW}:\right.\\
			&\quad\left.p_{XYZ}={}q_X\:q_{YZ|X},
		\right\}
	\end{align*}
	Then, $\mathcal{S}\subseteq\mathcal{R}$.
\end{theorem}

Informally, we may think of the auxiliary variable $U$ as containing information intended for both $P_2$ and $P_3$, $V$ as containing information for $P_2$, and $W$ as containing information for $P_3$.

We prove Theorem \ref{th:ach} using the Output Statistics of Random Binning (OSRB) framework introduced by Yassaee et al. \cite{osrb}. See Section \ref{sec:ach} and Appendix \ref{sec:app_ach} for a proof.

The inner bound above translates to the following upper bound on the minimum communication rate.
\begin{corollary}
	\label{cor:ach}
	\begin{align*}
		R_{\mathrm{opt}}\le \inf_{p_{XYZUVW}\in\mathcal{D}} \max\left\{ I(V;W|U)+I(U,V,W;X)\;,\right.\\
		\Big.\frac{1}{2}\left[ I(V;W|U) + I(U,V,W;X) + I(U;X,Y,Z) \right] \Big\}.
	\end{align*}
\end{corollary}

\subsection{Lower Bound}
\begin{theorem}
	\label{th:conv}
	\begin{align*}
		R_{\mathrm{opt}}\ge{}& \min_{p_{U|XYZ}}\max\Big\{I(Y;Z|U)+I(U,Y,Z;X)\;,\Big. \\
		&\quad\frac{1}{2}\Big. \left( I(Y;Z|U) + I(U,Y,Z;X) + I(U;X,Y,Z) \right) \Big\},
	\end{align*}
	where the min is over $p_{U|XYZ}$ with alphabet size at most $\abs{\mathcal{U}}\le \abs{\mathcal{X}}\abs{\mathcal{Y}}\abs{\mathcal{Z}}+3$.
\end{theorem}
See Section \ref{sec:conv} and Appendix \ref{sec:app_conv} for a proof.

\section{Special Cases and An Example}

%%special cases and example
\subsection{Special Cases}
\subsubsection{Point-to-Point Channel}
If $\abs{\mathcal{Z}}=1$ (or, more generally $Z\indep (X,Y)$), $P_3$ does not need to interact with $P_1,P_2$ and this becomes a point-to-point channel synthesis problem studied by Cuff \cite{cuff}. Let $\mathcal{R}_{X,Y}\subseteq \mathbb{R}_{\ge0}^2$ be the rate region of $(M,\Theta_1)$ for this problem. Cuff characterized the rate region precisely as

\begin{theorem}
	\cite[Theorem II.1]{cuff}
	\begin{align}
		\mathcal{R}_{X,Y}={}&\left\{ (R,R_1) \in \mathbb{R}^2 : \exists p_{U|XY} \in \mathcal{D}_{X,Y}\ \mathrm{s.t.} \right.\nonumber\\
		&\quad R\ge I(X;U)\nonumber\\
		&\quad \!\!\left.R+R_1\ge I(X,Y;U)\right\}\label{eq:cuff}
	\end{align}
	where $\mathcal{D}_{X,Y}=\left\{p_{U|XY}:I(X;Y|U)=0\right\}$.
	Moreover, $\abs{\mathcal{U}} \le \abs{\mathcal{X}} \abs{\mathcal{Y}} + 1$.
	\label{th:cuff}
\end{theorem}

If we substitute $U=W=$ const. and $X,Y,V\indep Z$ into our inner bound (and ignore $Z$), we get
\begin{align*}
	\mathcal{S}_{X,Y}={}&\left\{(R,R_1,R_2)\in\mathbb{R}^3_{\ge0}:\ \exists p_{XYV}\in\mathcal{D}_{X,Y}\ \mathrm{s.t.} \right.\\
	&\quad R\ge I(V;X)\\
	&\quad\!\!\left.R+R_1\ge I(V;X,Y)\right\}
\end{align*}
which recovers achievability part of Theorem \ref{th:cuff}. 
In this case, our lower bound on $R_{\mathrm{opt}}$ is also tight. From Theorem \ref{th:conv},
\begin{align*}
	R_{\mathrm{opt}}^{X,Y}\ge{}&\min_{p_{U|XYZ}}\max\Big\{I(U,Y;X)\;,\Big.\\
	&\quad\Big.\frac{1}{2} \left[ I(U,Y;X) + I(U;X,Y,Z) \right] \Big\} \\
	\ge{}& \min_{p_{U|XY}}I(U,Y;X)\\
	\ge{}& I(X;Y)
\end{align*}
which matches the mutual information lower bound.
\subsubsection{No Input}
If $\abs{\mathcal{X}}=1$ (or, more generally, $X\indep (Y,Z)$), then our problem reduces to Kurri et al.'s problem \cite{gowtham}. They give the following inner bound on the achievable rate region $\mathcal{R}_{Y,Z}$:
\begin{theorem}
	\cite[Theorem 1]{gowtham}
	\label{th:gowtham}
	Define $\mathcal{S}_{Y,Z}$ as
	\begin{align*}
		\mathcal{S}_{Y,Z}={}&\begin{cases}
			(R,R_1,R_2)\in\mathbb{R}^3_{\ge0}:\ \exists p_{YZUVW} \in\mathcal{D}_{Y,Z}\ \mathrm{s.t.}\\
			R\ge I(V;W|U)\\
			2R\ge I(V;W|U)+I(U;Y,Z)\\
			R+R_1\ge I(U,V;Y,Z)\\
			R+R_2\ge I(U,W;Y,Z)\\
			R+R_1+R_2\ge I(V;W|U)+I(U,V,W;Y,Z)\\
			2R+R_1+R_2\ge I(V;W|U)+I(U,V,W;Y,Z)\\
			\qquad\qquad\qquad\quad\ +I(U;Y,Z)
		\end{cases}
	\end{align*}
	where
	\begin{align*}
		\mathcal{D}_{Y,Z}={}&\begin{cases}
			p_{YZUVW}=q_{YZ}p_{UVW|YZ}\ \mathrm{s.t.}\\
			Y-(U,V)-(U,W)-Z\ \mathrm{is\ a\ Markov\ chain}.
		\end{cases}
	\end{align*}
	Then, $\mathcal{S}_{Y,Z}\subseteq \mathcal{R}_{Y,Z}$.
\end{theorem}
Notice that the Markov chain condition above is equivalent to the condition that $p_{YZUVW}=p_{UVW}\,p_{Y|UV}\,p_{Z|UW}$.
Substituting $(V,W)\indep X$ into our inner bound (Theorem \ref{th:ach}) reduces $\mathcal{S}$ to $\mathcal{S}_{Y,Z}$. Thus, our result matches the above theorem. Theorem \ref{th:gowtham} implies the following upper bound on the minimum communication rate $R_{\mathrm{opt}}^{Y,Z}$
\begin{align*}
		R_{\mathrm{opt}}^{Y,Z}\le{}&\min_{p_{U,Y,Z}}\Big[\max\Big\{I(Y;Z|U)\;,\Big.\Big.\\
	\Big.\Big.&\frac{1}{2}\left[I(Y;Z|U) + I(U;Y,Z)\right]\Big\}\Big].
\end{align*}
This is tight, as can be seen using Theorem \ref{th:conv}
\begin{align*}
	R_{\mathrm{opt}}^{Y,Z}\ge{}&\min_{p_{U,Y,Z}}\Big[\max\Big\{I(Y;Z|U)\;,\Big.\Big.\\
	\Big.\Big.&\frac{1}{2}\left[I(Y;Z|U) + I(U;Y,Z)\right]\Big\}\Big].
\end{align*}
This matches Kurri et al.'s characterisation of the minimum communication rate \cite[Theorem 2]{gowtham}.

\subsubsection{$Y=Z$}\label{sec:y=z}
For this case we give the minimum communication rate as:
\begin{theorem}
	\begin{align}
		R_{\mathrm{opt}}^{Y=Z}=I(X;Y)+\frac{1}{2}H(Y|X)\label{eq:y=z}.
	\end{align}
	\label{th:y=z}
\end{theorem}
\begin{IEEEproof}
	Let $\mathcal{R}_{Y=Z}\subseteq\mathbb{R}_{\ge0}^3$ be the set of achievable rates. 
Substituting $Y=Z$ and $V=W=$ const. into our inner bound (Theorem \ref{th:ach}) gives
\begin{corollary}
	\label{cor:y=z}
	Define $\mathcal{S}_{Y=Z}$ as
	\begin{align*}
		\mathcal{S}_{Y=Z}={}&\begin{cases}
			(R,R_1,R_2)\in\mathbb{R}^3_{\ge0}:\ \exists p_{X,Y,U}\in\mathcal{D}_{Y=Z}\ \mathrm{s.t.}\\
			R\ge I(U;X)+\frac{1}{2}I(U;Y|X)\\
			R+R_1\ge I(U;X,Y)\\
			R+R_2\ge I(U;X,Y)\\
		\end{cases}
	\end{align*}
	where
	\begin{align*}
		\mathcal{D}_{Y=Z}={}&\left\{ p_{X,Y,U}= q_X \;p_{U|X}\; p_{Y|U}\right.\\
		&\left.\mathrm{s.t.}\  p_{X,Y}={}q_X\;q_{Y|X}\right\}
	\end{align*}
	Then, $\mathcal{S}_{Y=Z}\subseteq\mathcal{R}_{Y=Z}$.
\end{corollary}

Thus, the minimum communication rate required satisfies
\begin{align*}
	R_{\mathrm{opt}}^{Y=Z}\le\min_{p_{UXY}\in\mathcal{D}_{Y=Z}}\left[I(U;X)+\frac{1}{2}I(U;Y|X)\right].
\end{align*}
Substituting $U=Y$ gives
\begin{align*}
	R_{\mathrm{opt}}^{Y=Z}\le I(Y;X)+\frac{1}{2}H(Y|X).
\end{align*}
This is actually tight as can be seen by using the lower bound (Theorem \ref{th:conv}):
\begin{align*}
	R_{\mathrm{opt}}^{Y=Z}\ge{}& \min_{p_{U|XY}} \Big[ \max\Big\{ H(Y|U)+I(U,Y;X)\;, \Big.\Big.\\
	&\Big.\Big.\frac{1}{2} \left[ H(Y|U) + I(U,Y;X) + I(U;X,Y) \right] \Big\} \Big]\\
	={}&\min_{p_{U|XY}}\Big[\max\Big\{H(Y|U)+I(U,Y;X)\;, \Big.\Big.\\
	&\frac{1}{2}\left[H(Y|U)+I(U;X|Y)+I(X;Y)\right.\\
	&\Big.\Big.\left.+I(U;X|Y)+I(U;Y)\right]\Big\}\Big]\\
	\ge{}& \max\left\{I(X;Y)\;,\; \frac{1}{2}\left[I(X;Y)+H(Y)\right]\right\}\\
	={}& \max\left\{I(X;Y)\;,\; I(X;Y) + \frac{1}{2}H(Y|X)\right\}\\
	={}&I(X;Y)+\frac{H(Y|X)}{2}.
\end{align*}
\end{IEEEproof}

The achievability part of Theorem \ref{th:y=z} can be alternatively obtained from \cite[Theorem II.1]{cuff} (Theorem \ref{th:cuff}); also see, \cite[Theorem 2]{bennett}.
Consider the operating point $R=I(X;Y),R_1=H(Y|X)$ in (\ref{eq:cuff}) obtained by choosing $U=Y$.
At this operating point, the scheme does not require any private randomization at the output terminal. To employ this scheme for the $Y=Z$ case, it is therefore sufficient if $P_1,P_2,P_3$ have $nH(Y|X)$ bits of common randomness.
There is a way to convert shared randomness into the requisite common randomness if $P_1$ sends an additional common message of $nH(Y|X)/2$ bits: $P_1$ takes $nH(Y|X)/2$ bits from each of $\Theta_1$ and $\Theta_2$ and sends their XOR as part of the message. Then, $P_2$ can XOR $\Theta_1$ with the message to obtain $\Theta_2$ and $P_3$ can likewise obtain $\Theta_1$. This way, $nH(Y|X)$ bits of common randomness are generated; the same idea was also employed in~\cite{gowtham}. With this additional message, the overall communication rate is $I(X;Y) + H(Y|X)/2$. 

\subsection{An Example: Binary Erasure Broadcast Channel}
As an example consider a binary erasure broadcast channel with independent erasures: $\mathcal{X}=\{0,1\}$, $\mathcal{Y}=\mathcal{Z}=\{0,1,\perp\}$. The input distribution is $X\sim\mathrm{Bernoulli}(1/2)$, and the channel is defined as follows, where $q_1,q_2 \in (0,1)$:
\begin{align}
	(Y,Z)=\begin{cases}
		(X,X) & \mathrm{w.p.\ } (1-q_1)(1-q_2)\\
		(X,\perp) & \mathrm{w.p.\ } (1-q_1)q_2\\
		(\perp,X) & \mathrm{w.p.\ }q_1(1-q_2)\\
		(\perp,\perp) & \mathrm{w.p.\ } q_1q_2
	\end{cases}
	\label{eq:ibec1}
\end{align}
For this channel, our results show that $R_{\mathrm{opt}}=(1-q_1q_2)$.

First, we show the achievability. Let $V\sim\mathrm{Bernoulli}(q_1)$ and $W\sim\mathrm{Bernoulli}(q_2)$ be independent random variables independent of $X$. Then define
\begin{align*}
	U={}&\begin{cases}
		\mathrm{Bernoulli}(1/2) & \mathrm{if\ } V=W=1\\
		X & \mathrm{otherwise}
	\end{cases}\\
	Y={}& \begin{cases}
		U & \mathrm{if\ } V=0\\
		\perp & \mathrm{otherwise}
	\end{cases}\\
	Z={}& \begin{cases}
		U & \mathrm{if\ } W=0\\
		\perp & \mathrm{otherwise}
	\end{cases}
\end{align*}
It can be easily checked that $p_{XYZUVW}$ belongs to $\mathcal{D}$ in Theorem~\ref{th:ach}.
%satisfies the rethis satisfies the required Markov chain constraints and results in the correct distribution $p_{XYZ}$.
The rate region $\mathcal{S}$ in the theorem for this choice of $U,V,W$ is given by:
\begin{align*}
	R\ge{}& 1-q_1q_2\\
	2R\ge{}& 2(1-q_1q_2)\\
	R+R_1\ge{}& 1-q_1q_2+h(q_1)\\
	R+R_2\ge{}& 1-q_1q_2+h(q_2)\\
	R+R_1+R_2\ge{}& 1-q_1q_2+h(q_1)+h(q_2)\\
	R+\frac{1}{2}\left[R_1+R_2\right]\ge{}&1-q_1q_2+\frac{1}{2}\left[h(q_1)+h(q_2)\right]
\end{align*}
This gives an upper bound on the minimum communication rate
\begin{align*}
	R_{\mathrm{opt}}\le 1-q_1q_2.
\end{align*}

We also have a matching lower bound from Theorem \ref{th:conv} \footnote{The lower bound of $R_{opt}\ge I(X;Y,Z)$ also follows from \cite{bennett}}:
\begin{align*}
	R_{\mathrm{opt}}\ge{}& \min_{p_{U|XYZ}}I(Y;Z|X)+I(U,Y,Z;X)\\
	\ge{}& I(Y,Z;X) = 1-q_1q_2.
\end{align*}

For comparison, consider the communication rate required when there is {\em no} shared (or common) randomness. Then the minimum communication rate required is
$\min_{p_{U|XYZ}} I(X,Y,Z;U)$ where the minimum is over $p_{U|XYZ}$ such that $X,Y,Z$ are conditionally independent conditioned on $U$~\cite[Section III-A]{cuff}. For $q_1=q_2=1/2$, this rate can be shown to be 1 bit, i.e., the trivial scheme of sending the input to the output terminals is optimal in the absence of any shared/common randomness. (To see this, we note that just to synthesise the point-to-point binary erasure channel with erasure probability at most 1/2, a communication rate of 1 bit is needed when the terminals do not have common randomness~\cite[Section II-F]{cuff}). In contrast, with shared randomness, the communication rate is only 3/4 bits.

In Appendix \ref{sec:app_eg}, we discuss a more general example of the binary erasure broadcast channel.\\

%Due to space constraints, the detailed proofs are deferred to the appendix. Here we sketch the proofs of the inner bound and of the lower bound.

\section{Proofs}

%%proofs

Due to space constraints, the detailed proofs are deferred to the appendices. Here we sketch the proofs of the inner bound and of the lower bound.

\subsection{Proof Sketch of Inner Bound}\label{sec:ach}
The inner bound proof uses Yassaee et al.'s OSRB framework \cite{osrb} together with the XOR trick mentioned in section \ref{sec:y=z}, which was also used in \cite{gowtham}.
We split $M,\Theta_1,\Theta_2$ into two parts each - $M=(M^{(a)},M^{(b)}),\ \Theta_i=(\Theta_i^{(a)},\Theta_i^{(b)})$ for $i=1,2$. We set $M^{(a)} = \Theta_1^{(a)} \oplus \Theta_2^{(a)}$ (Here, $\oplus$ stands for bitwise XOR). $P_2$ can recover $\Theta_2^{(a)}$ as $M^{(a)}\oplus\Theta_1^{(a)}$. Similarly, $P_3$ can recover $\Theta_1^{(a)}$. Thus, $(\Theta_1^{(a)},\Theta_2^{(a)})$ is known to all parties and can be used as common randomness. The second parts $M^{(b)},\Theta_1^{(b)},\Theta_2^{(b)}$ are used according to the OSRB framework. To this end we assign the following binning indices:
\begin{itemize}
	\item[-] To each $u^n\in\ \mathcal{U}^n$ assign three bin indices $m,m_0,f$ uniformly and independently with rates $R_m,R_{m_0},R_f$ respectively
	\item[-] To each $(u^n,v^n)\in\ \mathcal{U}^n\times\mathcal{V}^n$ assign two bin indices $b_1,f_1$ uniformly and independently with rates $R_{b_1},R_{f_1}$ respectively
	\item[-] To each $(u^n,w^n)\in\ \mathcal{U}^n\times\mathcal{W}^n$ assign two bin indices $b_2,f_2$ uniformly and independently with rates $R_{b_2},R_{f_2}$ respectively.
\end{itemize}
$m_0=(\Theta_1^{(a)},\Theta_2^{(a)})$ is the common randomness generated as above. $m=M^{(b)}$, $b_1=\Theta_1^{(b)}$, $b_2=\Theta_2^{(b)}$ are the message and shared random variables of the OSRB scheme. $f,f_1,f_2$ are "extra" shared random variables used for ease of analysis. During the analysis, we assume $P_1$ has access to $f,f_1,f_2$; $P_2$ has access to $f,f_1$ and $P_3$ has access to $f,f_2$. In the end, they will be set to some fixed values.

The protocol followed by $P_1,P_2,P_3$ is as follows:
\begin{enumerate}
	\item $P_1$ observes $x^n,\theta_1=(\theta_1^{(a)},b_1),\theta_2=(\theta_2^{(a)},b_2)$. It calculates $m_0=\theta_1^{(a)}\oplus \theta_2^{(a)}$. It then generates $u^n,v^n,w^n$ according to the conditional distribution $p_{U^nV^nW^n|M_0F\Theta_1F_1\Theta_2F_2X^n}$ (see equation (\ref{eq:pmf2})). $P_1$ then sends $(m_0,m(u^n))$ to $P_2$ and $P_3$.
	\item $P_2$ uses a Slepian-Wolf decoder to estimate $u^n,v^n$ from $m,m_0,b_1,f,f_1 $ and generates $y^n$ according to $p_{Y|UV}$.
	\item $P_3$ uses a Slepian-Wolf decoder to estimate $u^n,w^n$ from $m,m_0,b_2,f,f_2 $ and generates $z^n$ according to $p_{Z|UW}$.
\end{enumerate}
Informally, we need binning rates to be low enough so that $f,f_1,f_2$ can be eliminated. At the same time the binning rates must be high enough to ensure reliable decoding by $P_2$ and $P_3$. Putting all the constraints together results gives the following (See Appendix \ref{sec:app_ach} for details):
	\begin{align}
	\begin{split}
		R_m >{}& I(V;W|U) + I(U,V,W;X)\\
		R_m + R_{m_0} >{}& I(U;X,Y,Z)\\
		R_m + R_{m_0} + R_{b_1} >{}& I(U,V;X,Y,Z)\\
		R_m + R_{m_0} + R_{b_2} >{}& I(U,W;X,Y,Z)\\
		R_m + R_{m_0} + R_{b_1} + R_{b_2} >{}& I(V,W|U) + I(U,V,W;X,Y,Z)
	\end{split}\label{eq:skch1}
\end{align}
together with non-negativity of the rates involved.
We are interested in the total rates of $M,\Theta_1,\Theta_2$, which are
\begin{align*}
	R ={}& R_{m_0}/2 + R_m\\
	R_1 ={}& R_{m_0}/2 + R_{b_1}\\
	R_2 ={}& R_{m_0}/2 + R_{b_2}
\end{align*}
respectively. Substituting this into (\ref{eq:skch1}) and eliminating $R_{m_0}$ gives the claimed rate region.

\subsection{Proof Sketch of Lower Bound}\label{sec:conv}
Suppose $P_1,P_2,P_3$ have a protocol to simulate the broadcast channel $q_{YZ|X}$ on the input $q_X$.

We first lower bound the communication rate as follows:
\begin{align*}
	nR\ge{}& H(M)\ge H(M|\Theta_2)\\
	\ge{}& H(M|\Theta_2) - H(M|\Theta_1,\Theta_2) + H(M|\Theta_1,\Theta_2)\\
	& - H(M|X^n,\Theta_1,\Theta_2) \\
	={}& I(M;\Theta_1|\Theta_2)+ I(M;X^n|\Theta_1,\Theta_2) \\
	={}& I(\Theta_1;\Theta_2|M) + I(\Theta_1;M) - \underset{=0\;(\because\Theta_1\indep\Theta_2)}{\underbrace{I(\Theta_1;\Theta_2)}}\\
	& + I(M;X^n|\Theta_1,\Theta_2)\\
	\ge{}& I(\Theta_1;\Theta_2|M) + I(M;X^n|\Theta_1,\Theta_2)\\
	={}& I(\Theta_1;\Theta_2|M)+I(M,\Theta_1,\Theta_2;X^n)
\end{align*}

Now since, $(Y^n,Z^n)$ is conditionally independent of $X^n$ conditioned on $(M,\Theta_1,\Theta_2)$, we have $I(M,\Theta_1,\Theta_2;X^n)=I(M,\Theta_1,\Theta_2,Y^n,Z^n;X^n)$.
Furthermore, $Y^n$ is conditionally independent of $(\Theta_2,Z^n)$ conditioned on $(M,\Theta_1)$; and $Z^n$ is conditionally independent of $(\Theta_1,Y^n)$ conditioned on $(M,\Theta_2)$. Thus, $I(\Theta_1;\Theta_2|M)=I(\Theta_1,Y^n;\Theta_2,Z^n|M)$. Therefore, we have
\begin{align*}
	nR\ge{}& I(M,\Theta_1,\Theta_2,Y^n,Z^n;X^n) + I(\Theta_1,Y^n;\Theta_2,Z^n|M)\\
	\ge{}& I(Y^n;Z^n|M)+I(M,Y^n,Z^n;X^n)
\end{align*}
Using this we show
\begin{align}
	R\ge{}&\frac{1}{n}\sum_{i=1}^nI(Y_i;Z_i|M,Y^{i-1},Z^{i-1})\nonumber\\
	&\quad+\frac{1}{n}\sum_{i=1}^{n}I(M,Y^{i-1},Z^{i-1},Y_i,Z_i;X_i)\label{eq:conv1}
\end{align}
With similar arguments, we also lower bound the communication rate as
\begin{align}
	R\ge{}& \frac{1}{n}\sum_{i=1}^{n} \left[ I(M,Y^{i-1},Z^{i-1};X_i,Y_i,Z_i) \right.\nonumber\\
	& \quad\left. - 2I(X^{i-1},Y^{i-1},Z^{i-1};X_i,Y_i,Z_i)\right]\label{eq:conv2}
\end{align}
Let $T\sim\mathrm{Unif}\left([n]\right)$ be independent of all other random variables introduced so far and define $U = (M, Y^{T-1}, Z^{T-1}, T)$. Then equations (\ref{eq:conv1}) and (\ref{eq:conv2}) become
\begin{align}
	R\ge{}& I(Y_T;Z_T|U) + I(U,Y_T,Z_T;X_T)\label{eq:conv3}\\
	R\ge{}& I(U;X_T,Y_T,Z_T)\nonumber\\
	&-2I(X^{T-1},Y^{T-1},Z^{T-1};X_T,Y_T,Z_T)\label{eq:conv4}
\end{align}
We show that, without loss of generality, we can take the support of  $U$ to be of bounded size: $\abs{\mathcal{U}}\le \abs{\mathcal{X}}\abs{\mathcal{Y}}\abs{\mathcal{Z}}+3$.

Since $X,Y,Z$ are nearly i.i.d., the last term goes to 0 as $n\rightarrow\infty$ and $p_{X^nY^nZ^n}\rightarrow\prod q_Xq_{YZ|X}$. Using the fact that the cardinalities of all random variables involved are fixed independent of $n$, we show using continuity arguments that
\begin{align}
	R_{\mathrm{opt}}\ge{}&\min_{p_{U|XYZ}}\max\Big\{I(Y;Z|U)+I(U,Y,Z;X)\;,\Big. \nonumber\\
	&\Big.\qquad I(U;X,Y,Z)\Big\}\label{eq:conv5}
\end{align}
This clearly implies the statement of Theorem \ref{th:conv}. Although this statement is formally stronger than the statement of Theorem \ref{th:conv}, the two are actually equivalent. See Appendix \ref{sec:app_equiv} for a proof.

\IEEEtriggeratref{12}
\balance
\printbibliography

\clearpage
\appendices

\section{Proof of Theorem \ref{th:ach}}\label{sec:app_ach}
The proof uses Yassaee et. al.'s OSRB framework \cite{osrb} which works like so:\\
Suppose $X^n_{[m]},Z^n$ are random variables distributed according to $p_{X_{[m]}Z}$ i.i.d. We randomly bin the sequences of $X_{[m]}$. Specifically, a random binning is given by a collection of random maps $\{ \mathcal{B}_i: \mathcal{X}_i^n \rightarrow [2^{nR_i}] \}_{i\in[m]}$ where each sequence in $\mathcal{X}_i^n$ is mapped independently and uniformly to $[2^{nR_i}]$. This induces a random p.m.f. on the set $\mathcal{X}_{[m]}^n\times\mathcal{Z}\times\prod_{i=1}^{m}[2^{nR_i}]$, given by
\begin{align*}
	P(x_{[m]}^n,z^n,b_{[m]})= p_{X_{[m]}^n,Z^n}(x_{[m]},z^n) \prod_{i=1}^{m} \mathbf{1}_{\{\mathcal{B}_i(x_i^n)=b_i\}}
\end{align*}
Let $B_i$ denote the random variable $\mathcal{B}_i(X_i^n)$. It is clear that in expectation, the binning indices $B_{[m]}$ are distributed uniformly and independently from $Z^n$:
\begin{align*}
	\mathbb{E}_\mathcal{B}\left[P(z^n,b_{[m]})\right]=p_{Z^n}(z^n)\prod_{i=1}^{m}\frac{1}{2^{nR_i}}
\end{align*}
Yassaee et.al. give a condition on the rate tuple $R_{[m]}$ for this to hold with high probability:
\begin{lemma}
	\cite[Theorem 1]{osrb}\\
	\label{lm:osrb}
	If for every subset $S\subseteq[m]$,
	\begin{align*}
		\sum_{i\in S}R_i<H(X_{S}|Z)
	\end{align*}
	then,
	\begin{align*}
		\lim_{n\rightarrow\infty}\mathbb{E}_\mathcal{B}\norm{P(z^n,b_{[m]})-p_{Z^n}(z^n)\prod_{i=1}^m\frac{1}{2^{nR_i}}}_{\mathrm{TV}}=0
	\end{align*}
\end{lemma}
We are now ready to prove our inner bound
\begin{IEEEproof}[Proof of Theorem \ref{th:ach}]
	We split $M,\Theta_1,\Theta_2$ into two parts each - $M=(M^{(a)},M{(b)}),\ \Theta_i=(\Theta_i^{(a)},\Theta_i^{(b)})$ for $i=1,2$. We set $M^{(a)} = \Theta_1^{(a)} \oplus \Theta_2^{(a)}$ (Here, $\oplus$ stands for bitwise XOR). $P_2$ can recover $\Theta_2^{(a)}$ as $M^{(a)}\oplus\Theta_1^{(a)}$. Similarly, $P_3$ can recover $\Theta_1^{(a)}$. Thus, $(\Theta_1^{(a)},\Theta_2^{(a)})$ is known to all parties and can be used as common randomness. The second parts $M^{(b)},\Theta_1^{(b)},\Theta_2^{(b)}$ are used according to the OSRB framework. The overall strategy is as follows:
	
	Let $U,V,W$ be auxiliaries. Take a joint distribution of the form $p^{}_{XYZUVW}= q^{}_X\,p^{}_{UVW|X}\,p^{}_{Y|UV}\,p^{}_{Z|UW}$ such that $p^{}_{XYZ}=q^{}_X\,q^{}_{YZ|X}$, and assign random bin indices like so:
	\begin{itemize}
		\item[-] To each $u^n\in\ \mathcal{U}^n$ assign three bin indices $m,m_0,f$ uniformly and independently with rates $R_m,R_{m_0},R_f$ respectively
		\item[-] To each $(u^n,v^n)\in\ \mathcal{U}^n\times\mathcal{V}^n$ assign two bin indices $b_1,f_1$ uniformly and independently with rates $R_{b_1},R_{f_1}$ respectively
		\item[-] To each $(u^n,w^n)\in\ \mathcal{U}^n\times\mathcal{W}^n$ assign two bin indices $b_2,f_2$ uniformly and independently with rates $R_{b_2},R_{f_2}$ respectively.
	\end{itemize}
	$m_0=(\Theta_1^{(a)},\Theta_2^{(a)})$ is the common randomness generated as above. $m=M^{(b)}$, $b_1=\Theta_1^{(b)}$, $b_2=\Theta_2^{(b)}$ are the message and shared random variables of the OSRB scheme. $f,f_1,f_2$ are "extra" shared random variables used for ease of analysis. In the end they can be set to some fixed value without compromising the protocol.
	
	We also use Slepian-Wolf decoders to estimate $(u^n,v^n)$ from $(m,m_0,b_1,f,f_1)$. A Slepian-Wolf decoder will pick the unique sequence $(\hat{u}_{(1)}{}^n,\hat{v}^n)$ which is jointly typical with the bin indices $(m,m_0,b_1,f,f_1)$, if such a sequence exists. If no such sequence exists or it is not unique, the decoder outputs an arbitrary fixed sequence. \\
	Similarly we use another Slepian-Wolf decoder to obtain estimates $(\hat{u}_{(2)}{}^n,\hat{w}^n)$ of $(u^n,w^n)$ from $(m,m_0,b_2,f,f_2)$.\\
	The resulting random p.m.f. is
	\begin{align}
		&P\left(x^n,y^n,z^n,u^n,v^n,w^n,m,m_0,b_1,b_2, \right.\nonumber\\
		&\qquad\qquad\qquad\left.f,f_1,f_2,\hat{u}_{(1)}^n,\hat{v}^n,\hat{u}_{(2)}^n,\hat{w}^n\right)\nonumber\\
		={}&q(x^n) p(u^n,v^n,w^n|x^n) p(y^n|u^n,v^n) p(z^n|u^n,w^n)\nonumber\\
		&\times P(m|u^n) P(m_0,f|u^n)\nonumber\\
		&\times P(b_1,f_1|u^n,v^n) P(b_2,f_2|u^n,v^n)\nonumber\\
		&\times P^{\mathrm{SW}} \left( \hat{u}_{(1)}^n,\hat{v}^n|m,m_0,b_1,f_1,f\right)\nonumber\\
		&\times P^{\mathrm{SW}} \left( \hat{u}_{(2)}^n,\hat{w}^n|m,m_0,b_2,f_2,f\right)\label{eq:pmf1}\\
		={}& q(x^n)P(m_0,f,b_1,f_1,b_2,f_2,u^n,v^n,w^n|x^n) \nonumber\\
		&\times P(m|u^n) P^{\mathrm{SW}} \left( \hat{u}_{(1)}^n,\hat{v}^n|m,m_0,b_1,f_1,f\right) \nonumber\\
		&\times P^{\mathrm{SW}} \left( \hat{u}_{(2)}^n, \hat{w}^n | m ,m_0, b_2,f_2,f\right)\nonumber\\
		={}&q(x^n)P(m_0,f,b_1,f_1,b_2,f_2|x^n)\nonumber\\
		&\times p(u^n,v^n,w^n|m_0,f,b_1,f_1,b_2,f_2,x^n) P(m|u^n)\nonumber\\
		&\times P^{\mathrm{SW}} \left( \hat{u}_{(1)}^n,\hat{v}^n|m,m_0,b_1,f_1,f\right) \nonumber\\
		&\times P^{\mathrm{SW}} \left( \hat{u}_{(2)}^n, \hat{w}^n|m,m_0,b_2,f_2,f \right) \label{eq:pmf2}
	\end{align}
	The protocol followed by $P_1,P_2,P_3$ is (assume that $f$ is known to all parties; $f_1$ is known to $P_1$ and $P_2$; $f_2$ is known to $P_1$ and $P_3$. We will justify this later.):
	\begin{itemize}
		\item $P_1$ chooses $m,u^n,v^n,w^n$ according to $p(u^n,v^n,w^n|m_0,f,b_1,f_1,b_2,f_2,x^n)P(m|u^n)$ from equation (\ref{eq:pmf2}) and sends $(m_0,m)$ as the common message.
		\item $P_2$ uses a Slepian-Wolf decoder to obtain estimates $(\hat{u}_{(1)}{}^n,\hat{v}^n)$ of $(u^n,v^n)$ from $(m,m_0,b_1,f,f_1) $.
		\item $P_3$ uses a Slepian-Wolf decoder to obtain estimates $(\hat{u}_{(2)}{}^n,\hat{w}^n)$ of $(u^n,w^n)$ from $(m,m_0,b_2,f,f_2) $.
		\item $P_2$ and $P_3$ generate $y^n$ and $z^n$ according to $\prod_{i=1}^{n} p_{Y|U,V} (\cdot|\left(\hat{u}_{(1)}\right){}_i,\hat{v}_i)$ and $\prod_{i=1}^{n} p_{Z|U,W}(\cdot|\left(\hat{u}_{(2)}\right){}_i,\hat{w}_i)$ respectively.
	\end{itemize}
	
	In order for this protocol to succeed, the following constraints must be met:\\
	\textbf{Uniformity conditions:} We need the bin indices $m_0,b_1,b_2,f,f_1,f_2$ to be jointly uniform and independent of $X^n$. This is because $\Theta_1,\Theta_2,X^n$ are independent, and we want to interpret $\left(\Theta_1^{(a)},\Theta_2^{(a)}\right),\Theta_1^{(b)},\Theta_2^{(b)}$ as $m_0,b_1,b_2$. Furthermore, we want to eventually fix a particular value of $(f,f_1,f_2)$, so we need them to be uniform and independent of $m_0,b_1,b_2,X^n$.\\
	Applying lemma \ref{lm:osrb} with $m=3$, $X_{[m]}=\left(U,\left(U,V\right),\left(U,W\right)\right)$, and $b_{[m]}=\left(\left(m_0,f\right),\left(b_1,f_1\right),\left(b_2,f_2\right)\right)$ and $Z=X$ and eliminating redundant inequalities, we get
	\begin{align}
		\begin{split}
			R_{m_0}+R_f<{}&H(U|X)\\
			R_{m_0}+R_f+R_{b_1}+R_{f_1}<{}& H(U,V|X)\\
			R_{m_0}+R_f+R_{b_2}+R_{f_2}<{}& H(U,W|X)\\
			R_{m_0}+R_f\quad\ \;&\\
			+R_{b_1}+R_{f_1}+R_{b_2}+R_{f_2}<{}& H(U,V,W|X)
		\end{split}\label{eq:unif1}
	\end{align}
	If these are satisfied, $q(x^n)P(m_0,f,b_1,f_1,b_2,f_2|x^n)$ will tend to $q(x^n) p^{\mathrm{Unif}}(m_0) p^{\mathrm{Unif}}(f) p^{\mathrm{Unif}}(b_1) p^{\mathrm{Unif}}(f_1)\\ p^{\mathrm{Unif}}(b_2) p^{\mathrm{Unif}}(f_2)$ in total variation distance as $n\rightarrow\infty$.
	
	Next, in order to eliminate $f,f_1,f_2$, we further need $f,f_1,f_2$ to be independent of $X,Y,Z$. Applying lemma \ref{lm:osrb} with $m=3$, $X_{[m]}=\left(U,\left(U,V\right),\left(U,W\right)\right)$,  $b_{[m]}=\left(f, f_1, f_2\right)$, and $Z=(X,Y,Z)$, we get (after eliminating redundant inequalities):
	\begin{align}
		\begin{split}
			R_f <{}& H(U|X,Y,Z)\\
			R_f+R_{f_1} <{}& H(U,V|X,Y,Z)\\
			R_f+R_{f_2} <{}& H(U,W|X,Y,Z)\\
			R_f+R_{f_1}+R_{f_2} <{}& H(U,V,W|X,Y,Z)
		\end{split}\label{eq:unif2}
	\end{align}
	As long as these are satisfied, the marginal distribution $P(f,f_1,f_2,x^n,y^n,z^n)$ will tend to $P^{\mathrm{Unif}}(f,f_1,f_2)p(x^n,y^n,z^n)$ as $n\rightarrow\infty$. Thus, $P_1,P_2,P_3$ can agree beforehand on a particular value of $f,f_1,f_2$ without affecting the joint distribution of $X^n,Y^n,Z^n$.\\\\
	\textbf{Slepian-Wolf conditions:}
	These are required for the Slepian Wolf decoders used by $P_2$ and $P_3$ to succeed with high probability \cite{SW}:
	\begin{align*}
		\begin{split}
			R_{b_1}+R_{f_1} \ge{}& H(V|U)\\
			R_m + R_{m_0} + R_f + R_{b_1} + R_{f_1} \ge{}& H(U,V)\\
			R_{b_2}+R_{f_2} \ge{}& H(W|U)\\
			R_m + R_{m_0} + R_f + R_{b_2} + R_{f_2} \ge{}& H(U,W)\\
		\end{split}
	\end{align*}
	We can reduce these 4 inequalities to 3 (i.e. the following inequalities imply the above inequalities):
	\begin{align}
		\begin{split}
			R_{b_1}+R_{f_1} \ge{}& H(V|U)\\
			R_{b_2}+R_{f_2} \ge{}& H(W|U)\\
			R_m + R_{m_0} + R_f \ge{}& H(U)\\
		\end{split}\label{eq:unif3}
	\end{align}
	If these are satisfied, the Slepian-Wolf distributions $P^{\mathrm{SW}} \left( \hat{u}_{(1)}^n,\hat{v}^n|m,m_0,b_1,f_1,f\right)$ and $ P^{\mathrm{SW}} \left( \hat{u}_{(2)}^n,\hat{w}^n|m,m_0,b_2,f_2,f\right)$ will tend to $\mathbf{1}_{\{\hat{u}_{(1)}^n=u^n\}}\mathbf{1}_{\{\hat{u}_{(2)}^n=u^n\}}\mathbf{1}_{\{\hat{v}^n=v^n\}}\mathbf{1}_{\{\hat{w}^n=w^n\}}$ as $n\rightarrow\infty$.\\\\
	\textbf{Putting it all together}:\\
	We can assume the Slepian-Wolf conditions (\ref{eq:unif3}) are tight since, if (\ref{eq:unif3}) are satisfied, we can reduce the rates $R_f,R_{f_1},R_{f_2}$ without affecting (\ref{eq:unif1}) or (\ref{eq:unif2}). We now replace (\ref{eq:unif3}) with equalities, solve for $R_f,R_{f_1},R_{f_2}$ and substitute into (\ref{eq:unif1}) and (\ref{eq:unif2}). This gives (after eliminating redundant inequalities):
	\begin{align}
		\begin{split}
			R_m >{}& I(V;W|U)\\
			& + I(U,V,W;X)\\
			R_m + R_{m_0} >{}& I(U;X,Y,Z)\\
			R_m + R_{m_0} + R_{b_1} >{}& I(U,V;X,Y,Z)\\
			R_m + R_{m_0} + R_{b_2} >{}& I(U,W;X,Y,Z)\\
			R_m + R_{m_0} + R_{b_1} + R_{b_2} >{}& I(V,W|U)\\
			& + I(U,V,W;X,Y,Z)
		\end{split}\label{eq:res1}
	\end{align}
	The rates we care about are those of $M,\Theta_1,\Theta_2$. These are, respectively:
	\begin{align*}
		R ={}& \frac{R_{m_0}}{2} + R_m\\
		R_1 ={}& \frac{R_{m_0}}{2} + R_{b_1}\\
		R_2 ={}& \frac{R_{m_0}}{2} + R_{b_2}
	\end{align*}
	Substituting into (\ref{eq:res1}) gives:
	\begin{align*}
		R - \frac{R_{m_0}}{2} >{}& I(V;W|U) + I(U,V,W;X)\\
		R + \frac{R_{m_0}}{2} >{}& I(U;X,Y,Z)\\
		R + R_1 >{}& I(U,V;X,Y,Z)\\
		R + R_2 >{}& I(U,W;X,Y,Z)\\
		R + R_1 + R_2 - \frac{R_{m_0}}{2} >{}& I(V;W|U) + I(U,V,W;X,Y,Z)
	\end{align*}
	We just need to eliminate $R_{m_0}$ from the above conditions together with $R_{m_0}\ge 0$. This gives:
	\begin{align*}
		R >{}& I(V;W|U) + I(U,V,W;X)\\
		2R >{}& I(V;W|U)+ I(U,V,W;X)\\
		& + I(U;X,Y,Z)\\
		R + R_1 >{}& I(U,V;X,Y,Z)\\
		R + R_2 >{}& I(U,W;X,Y,Z)\\
		R + R_1 + R_2 >{}& I(V;W|U) + I(U,V,W;X,Y,Z)\\
		2R + R_1 + R_2 >{}& I(V;W|U) + I(U,V,W;X,Y,Z)\\
		& + I(U;X,Y,Z)
	\end{align*}
	which completes the proof.
\end{IEEEproof}
\section{Proof of Lower Bound}
\label{sec:app_conv}
As a first step, we prove the following bound on the cardinality of the auxiliary variable, based on Caratheodory's Theorem.
\begin{lemma}[Cardinality Bound]
	\label{lm:card_bound}
	Given any random variables $X,Y,Z,W$ distributed according to $p_{XYZW}$, there exists a distribution $q_{XYZU}$ such that
	\begin{align*}
		\abs{\mathcal{U}}\le{}& \abs{\mathcal{X}} \abs{\mathcal{Y}} \abs{\mathcal{Z}} + 3\\
		p_{XYZ}={}&q_{XYZ}\\
		I_p(Y;Z|W)={}& I_q(Y;Z|U)\\
		I_p(W,Y,Z;X)={}& I_q(U,Y,Z;X)\\
		I_p(W;X,Y,Z)={}& I_q(U;X,Y,Z)
	\end{align*}
\end{lemma}
\begin{IEEEproof}
	Consider the set $\mathcal{A}\subseteq \mathbb{R}^{\abs{\mathcal{X}} \abs{\mathcal{Y}} \abs{\mathcal{Z}} + 4}$ where the first $\abs{\mathcal{X}} \abs{\mathcal{Y}} \abs{\mathcal{Z}}$ coordinates are probability masses of a distribution $\pi_{XYZ}$ and the last 4 coordinates are the corresponding (conditional) entropies $H(X,Y,Z), H(X|Y,Z), H(Y|Z)$ and $H(Y)$. Note that $\mathcal{A}$ is connected and compact, since it is a continuous image of the probability simplex $\Delta^{\abs{\mathcal{X}} \abs{\mathcal{Y}} \abs{\mathcal{Z}}}$.
	
	Consider the point $a\in\mathcal{R}^{\abs{\mathcal{X}} \abs{\mathcal{Y}} \abs{\mathcal{Z}} + 4}$, whose first $\abs{\mathcal{X}} \abs{\mathcal{Y}} \abs{\mathcal{Z}}$ coordinates are $p_{XYZ}$ and the last 4 coordinates are $H(X,Y,Z|W), H(X|Y,Z,W), H(Y|Z,W), H(Y|W)$.
	Since $p_{XYZ}=\sum_{w\in\mathcal{W}}p_W(w)p_{XYZ|W=w}$, the point $a$ lies in the convex hull of $\mathcal{A}$.
	Also, $\mathcal{A}$ lies inside a codimension 1 subspace of $\mathbb{R}^{\abs{\mathcal{X}} \abs{\mathcal{Y}} \abs{\mathcal{Z}} + 4}$, since the sum of probabilities needs to be 1.
	
	Therefore, by Caratheodory's theorem, $a$ is a convex combination of $\abs{\mathcal{X}} \abs{\mathcal{Y}} \abs{\mathcal{Z}} + 3$ points in $\mathcal{A}$. We will use these points to construct $q_{XYZU}$.
	Label the points $u=1,2,\ldots,\abs{\mathcal{X}} \abs{\mathcal{Y}} \abs{\mathcal{Z}} + 3$. The weights of the points represent $q_U(u)$ and the first $\abs{\mathcal{X}} \abs{\mathcal{Y}} \abs{\mathcal{Z}} $ coordinates represent $q_{XYZ|U=u}$.
	By construction of the point $a$, we get the following
	\begin{align*}
		p_{XYZ}={}&q_{XYZ}\\
		H_p(Y|W)={}&H_q(Y|U)\\
		H_p(Y|Z,W)={}& H_q(Y|Z,U)\\
		H_p(X|Y,Z,W)={}& H_q(X|Y,Z,U)\\
		H_p(X,Y,Z|W)={}& H_q(X,Y,Z|U)
	\end{align*}
	In addition to these four conditional entropy equalities, we also know that $H_p(X)=H_q(X)$ and $H_p(X,Y,Z)=H_q(X,Y,Z)$ since $p_{XYZ}=q_{XYZ}$. With this, we can obtain the promised mutual information equalities as follows:
	\begin{align*}
		I_q(Y;Z|U)={}& H_q(Y|U)-H_q(Y|Z,U)\\
		={}&H_p(Y|W)-H_p(Y|Z,W)= I_p(Y;Z|W)\\
		I_q(U,Y,Z;X)={}&H_q(X)-H(X|U,Y,Z)\\
		={}&H_p(X)-H_p(X|W,Y,Z)= I_p(W,Y,Z;X)\\
		I_q(U;X,Y,Z)={}&H_q(X,Y,Z)-H_q(X,Y,Z|U)\\
		={}&H_p(X,Y,Z)-H_p(X,Y,Z|W)\\
		={}& I_p(W;X,Y,Z)
	\end{align*}
\end{IEEEproof}
\begin{IEEEproof}[Proof of Theorem \ref{th:conv}]
	Suppose $P_1,P_2,P_3$ have a protocol to simulate the broadcast channel $q_{YZ|X}$ on the input $q_X$. That is, there is an encoder $\mathrm{Enc}:\mathcal{X}^n\times\left[2^{nR_1}\right]\times\left[2^{nR_2}\right]\rightarrow\left[2^{nR}\right]$ which $P_1$ uses to map $X^n,\Theta_1,\Theta_2$ to the message $M$. And, there are decoders $\mathrm{Dec}_1:\left[2^{nR}\right] \times\left[2^{nR_1}\right] \times\mathcal{Y}^n$ and $\mathrm{Dec}_2:\left[2^{nR}\right] \times\left[2^{nR_2}\right] \times\mathcal{Z}^n$ which are used by $P_2$ and $P_3$ to map $M,\Theta_1$ and $M,\Theta_2$ to $Y^n$ and $Z^n$ respectively.\\
	We can lower bound $R$ by
	\begin{align*}
		nR\ge{}& H(M)\ge H(M|\Theta_2)\\
		\ge{}& H(M|\Theta_2) - H(M|\Theta_1,\Theta_2) + H(M|\Theta_1,\Theta_2)\\
		& - H(M|X^n,\Theta_1,\Theta_2) \\
		={}& I(M;\Theta_1|\Theta_2)+ I(M;X^n|\Theta_1,\Theta_2) \\
		={}& I(\Theta_1;\Theta_2|M) + I(\Theta_1;M) - \underset{=0\;(\because\Theta_1\indep\Theta_2)}{\underbrace{I(\Theta_1;\Theta_2)}}\\
		& + I(M;X^n|\Theta_1,\Theta_2)\\
		\ge{}& I(\Theta_1;\Theta_2|M) + I(M,\Theta_1,\Theta_2;X^n)
	\end{align*}
	Now since, conditioned on $(M,\Theta_1,\Theta_2)$, $(Y^n,Z^n)$ is a independent of $X^n$, we have $I(M,\Theta_1,\Theta_2;X^n)=I(M,\Theta_1,\Theta_2,Y^n,Z^n;X^n)$.
	Furthermore, conditioned on $(M,\Theta_1)$, $Y^n$ is independent of $\Theta_2$ and, conditioned on $(M,\Theta_2)$, $Z^n$ is independent of $\Theta_1$. Thus, $I(\Theta_1;\Theta_2|M)=I(\Theta_1,Y^n;\Theta_2,Z^n|M)$. Therefore, we have
	\begin{align*}
		nR\ge{}& I(\Theta_1,Y^n;\Theta_2,Z^n|M) + I(M,\Theta_1,\Theta_2,Y^n,Z^n;X^n)\\
		\ge{}& I(Y^n;Z^n|M)+I(M,\Theta_1,\Theta_2Y^n,Z^n;X^n)\\
		\ge{}& I(Y^n;Z^n|M) + I(M,Y^n,Z^n;X^n)\\
		={}& \sum_{i=1}^{n} \left[I(Y_i;Z^n|M,Y^{i-1}) + I(M,Y^n,Z^n;X_i|X^{i-1})\right]\\
		={}& \sum_{i=1}^{n} \left[I(Y_i;Z^n|M,Y^{i-1}) + I(M,Y^n,Z^n,X^{i-1};X_i)\right.\\
		&\qquad\left.-I(X_i;X^{i-1})\right]\\
		\ge{}& \sum_{i=1}^{n} \left[I(Y_i;Z^i|M,Y^{i-1}) + I(M,Y^i,Z^i;X_i)\right]\\
		\ge{}& \sum_{i=1}^{n} \left[I(Y_i;Z_i|M,Y^{i-1},Z^{i-1})\right.\\
		&\qquad\left. + I(M,Y^{i-1},Z^{i-1},Y_i,Z_i;X_i)\right]\\
		\therefore R\ge{}& \frac{1}{n}\sum_{i=1}^{n}I(Y_i;Z_i|M,Y^{i-1},Z^{i-1})\\
		& + \frac{1}{n} \sum_{i=1}^{n} I(M,Y^{i-1},Z^{i-1},Y_i,Z_i;X_i)
	\end{align*}
	Let $T$ be a random variable distributed uniformly on $[n]$ and independent of all $X^n,Y^n,Z^n,M,\Theta_1,\Theta_2$.\\
	Then,
	\begin{align*}
		\frac{1}{n}\sum_{i=1}^nI(Y_i;Z_i|M,Y^{i-1},Z^{i-1}) ={}& I(Y_T;Z_T|M,Y^{T-1},Z^{T-1},T)
	\end{align*}
	and
	\begin{align*}
		&\frac{1}{n}\sum_{i=1}^nI(M,Y^{i-1},Z^{i-1},Y_i,Z_i;X_i)\\
		={}& I(M,Y^{T-1},Z^{T-1},Y_T,Z_T;X_T|T)\\
		={}& I(M,Y^{T-1},Z^{T-1},T,Y_T,Z_T;X_T) - I(T;X_T)\\
		={}& I(M,Y^{T-1},Z^{T-1},T,Y_T,Z_T;X_T)
	\end{align*}
	where the last step follows from $I(T;X_T)=0$ since $X^n$ is i.i.d.\\
	Define $U=(M,Y^{T-1},Z^{T-1},T)$. Then we get the bound
	\begin{align}
		R\ge{}& I(Y_T;Z_T|U)+I(U,Y_T,Z_T;X_T) \label{eq:1}
	\end{align}
	We can also bound $R$ another way:
	\begin{align}
		nR\ge{}& H(M)\nonumber\\
		\ge{}&  I(M;X^n,Y^n,Z^n)\nonumber\\
		\ge{}& \sum_{i=1}^{n} I(M;X_i,Y^n,Z^n|X^{i-1})\nonumber\\
		\ge{}& \sum_{i=1}^{n} I(M;X_i,Y^i,Z^i|X^{i-1})\nonumber\\
		\ge{}& \sum_{i=1}^{n}  I(M;X_i,Y_i,Z_i| X^{i-1},Y^{i-1},Z^{i-1}) \nonumber\\
		\ge{}& \sum_{i=1}^{n} \left[ I(M;X_i,Y_i,Z_i|Y^{i-1},Z^{i-1})\right.\nonumber\\
		& \qquad\left.- I(X^{i-1};X_i,Y_i,Z_i| Y^{i-1},Z^{i-1})\right]\nonumber\\
		={}& \sum_{i=1}^{n} \left[ I(M,Y^{i-1},Z^{i-1};X_i,Y_i,Z_i) \right.\nonumber\\
		& \qquad- I(X^{i-1};X_i,Y_i,Z_i|Y^{i-1},Z^{i-1})\nonumber\\
		&\qquad\left. - I(Y^{i-1},Z^{i-1};X_i,Y_i,Z_i) \right]\nonumber\\
		\ge{}& \sum_{i=1}^{n} \left[ I(M,Y^{i-1},Z^{i-1};X_i,Y_i,Z_i) \right.\nonumber\\
		& \qquad\left. - 2I(X^{i-1},Y^{i-1},Z^{i-1};X_i,Y_i,Z_i)\right]\label{eq:conv_res2}
	\end{align}
	The offending term $I(X^{i-1},Y^{i-1},Z^{i-1};X_i,Y_i,Z_i)$ tends to 0 as $n\rightarrow\infty$ due to $X^n,Y^n,Z^n$ being nearly i.i.d. To show this, we use a lemma due to Cuff which states
	\begin{lemma}
		\cite[Lemma VI.3]{cuff}\\
		\label{lm:2}
		Let $p_{W^n}$ be a nearly i.i.d. distribution, in the sense that, there exists a distribution $q_W$ such that
		\begin{align*}
			\norm{p_{W^n}-\prod_{i=1}^{n}q_W}_{\mathrm{TV}}<\epsilon
		\end{align*}
		Then,
		\begin{align*}
			\frac{1}{n}\sum_{i=1}^nI(W_i;W^{i-1})\le 4\epsilon\left(\log\abs{\mathcal{W}}+\log(\frac{1}{\epsilon})\right)
		\end{align*}
	\end{lemma}
	For our proof, we set $W^n=(X^n,Y^n,Z^n)$ which gives us
	\begin{align}
			&\frac{1}{n} \sum_{i=1}^{n} I(X_i,Y_i,Z_i;X^{i-1},Y^{i-1},Z^{i-1})\nonumber\\
			\le{}& 4\varepsilon\left( \log \abs{\mathcal{X}} \abs{\mathcal{Y}} \abs{\mathcal{Z}} + \log(\frac{1}{\varepsilon}) \right) =: g(\varepsilon) \label{eq:g}
	\end{align}
	Thus, our bound becomes
	\begin{align}
		R\ge{}&I(U;X_T,Y_T,Z_T)-2g(\varepsilon)  \label{eq:3}
	\end{align}
	By lemma \ref{lm:card_bound}, we can find a distribution $p_{XYZU}$ with $\abs{\mathcal{U}}\le \abs{\mathcal{X}}\abs{\mathcal{Y}}\abs{\mathcal{Z}}+3$ such that\\
	\begin{align*}
		p_{XYZ}={}&p_{X_TY_TZ_T}\\
		I_p(Y;Z|U)={}&I(Y_T,Z_T|U)\\
		I_p(U,Y,Z;X)={}&I(U,Y_T,Z_T;X_T)\\
		I_p(U;X,Y,Z)={}&I(U;X_T,Y_T,Z_T)
	\end{align*}
	
	So far, we have shown that for every $\varepsilon>0$, there exists a distribution $p_{UXYZ}\in \mathcal{D}_\varepsilon$ such that 
	\begin{align}
		R_{\mathrm{opt}}\ge{}& \max\{I(Y;Z|U)+I(U,Y,Z;X)\;,\nonumber\\
		&\qquad\ I(U;X,Y,Z) -2g(\varepsilon)\}\label{eq:conv_ge}
	\end{align}
	where
	\begin{align*}
		\mathcal{D}_\varepsilon=\left\{p_{XYZU}:	\norm{p_{XYZ}- q_Xq_{YZ|X}}_{\mathrm{TV}} < \varepsilon\right.\\ \left.\mathrm{and}\quad\abs{\mathcal{U}} \le \abs{\mathcal{X}}\abs{\mathcal{Y}}\abs{\mathcal{Z}} + 3\right\}
	\end{align*}
	We are almost done. We just need to let  $\varepsilon\rightarrow0$ and ensure there are no discontinuity issues. For this, define $\tilde{\mathcal{S}}_{\varepsilon,\delta}$ to be the region
	\begin{align*}
		\tilde{\mathcal{S}}_{\varepsilon,\delta} ={}& \left\{R\in\mathbb{R}:\exists p_{XYZU}\in\mathcal{D}_\varepsilon\ \mathrm{s.t.}\right.\\
		&\ \ R\ge \max\left\{I(Y;Z|U)+I(U,Y,Z;X)\;,\right.\\
		&\left.\left.\ I(U;X,Y,Z)-2\delta\right\}\right\}
	\end{align*}
	To complete the proof we need to show that $\bigcap_{\varepsilon>0}\tilde{\mathcal{S}}_{\varepsilon,g(\varepsilon)}=\tilde{\mathcal{S}}_{0,0}$.

	Note that $\mathcal{D}_\varepsilon$ is decreasing as $\varepsilon\rightarrow0$; it is compact due to the cardinality bound; and $\mathcal{\tilde{S}}_{\varepsilon,0}$ is a continuous image of $\mathcal{D}_\varepsilon$. Thus,
	\begin{align*}
		\bigcap_{\varepsilon>0}\mathcal{\tilde{S}}_{\varepsilon,0}=\mathcal{\tilde{S}}_{0,0}
	\end{align*}
	To complete the proof, we need to show that
	\begin{align*}
		\bigcap_{\varepsilon>0}\mathcal{\tilde{S}}_{\varepsilon,g(\varepsilon)}=\bigcap_{\varepsilon>0}\mathcal{\tilde{S}}_{\varepsilon,0}
	\end{align*}
	$\bigcap_{\varepsilon>0} \mathcal{\tilde{S}}_{\varepsilon,g(\varepsilon)} \supseteq \bigcap_{\varepsilon>0} \mathcal{\tilde{S}}_{\varepsilon,0}$ is clear.
	
	To show $\bigcap_{\varepsilon>0} \mathcal{\tilde{S}}_{\varepsilon,g(\varepsilon)} \subseteq \bigcap_{\varepsilon>0} \mathcal{\tilde{S}}_{\varepsilon,0}$, take any $r\in\mathbb{R}$ such that  $r\notin \bigcap_{\varepsilon>0} \mathcal{\tilde{S}}_{\varepsilon,0}$.
	Let $r'=\inf\left\{\bigcap_{\varepsilon>0} \mathcal{\tilde{S}}_{\varepsilon,0}\right\}$.
	Then, $\frac{r+r'}{2} \notin \bigcap_{\varepsilon>0} \mathcal{\tilde{S}}_{\varepsilon,0}$.
	Choose $\varepsilon$ small enough so that $\frac{r+r'}{2}\notin \mathcal{\tilde{S}}_{\varepsilon,0}$ and $2g(\varepsilon)<r'-r$.
	This is possible since $\lim_{\varepsilon\rightarrow0}g(\varepsilon)=0$. This implies $r\notin\bigcap_{\varepsilon>0}\mathcal{\tilde{S}}_{\varepsilon,g(\varepsilon)}$.
\end{IEEEproof}
Finally, we get the lower bound:
\begin{align*}
	R_{\mathrm{opt}}\ge{}& \min_{p_{U|X,Y,Z}}\Big[\max\Big\{I(Y;Z|U) + I(U,Y,Z;X)\;,\Big.\Big.\\
	&\left.\left. I(U;X,Y,Z)\right\}\right]\\
	\implies R_{\mathrm{opt}}\ge{}& \min_{p_{U|X,Y,Z}}\Big[\max\Big\{I(Y;Z|U) + I(U,Y,Z;X)\;,\Big.\Big.\\
	&\left.\left.\frac{I(Y;Z|U) + I(U,Y,Z;X) + I(U;X,Y,Z)}{2}\right\}\right]\\
\end{align*}
\section{Another Example}
\label{sec:app_eg}
Consider the following channel, where the input $X\sim \mathrm{Bernoulli}(1/2)$:\\
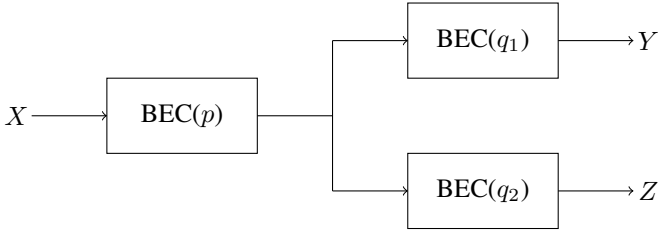
\begin{figure}[H]
	\begin{tikzpicture}[scale=2]
		\draw (.4,0) node{$X$};
		\draw[->] (.5,0) -- (1,0);
		\draw (1,-.25) rectangle (2,.25);
		\draw (1.5,0) node{BEC($p$)};
		\draw (2,0) -- (2.5,0);
		\draw (2.5,-.5) -- (2.5,.5);
		\draw[->] (2.5,.5) -- (3,.5);
		\draw[->] (2.5,-.5) -- (3,-.5);
		\draw (3,.25) rectangle (4,.75);
		\draw (3,-.25) rectangle (4,-.75);
		\draw (3.5,.5) node{BEC($q_1$)};
		\draw (3.5,-.5) node{BEC($q_2$)};
		\draw[->] (4,.5) -- (4.5,.5);
		\draw[->] (4,-.5) -- (4.5,-.5);
		\draw (4.6,.5) node{$Y$};
		\draw (4.6,-.5) node{$Z$};
	\end{tikzpicture}
\end{figure}
This broadcast channel has the conditional distribution $q_{YZ|X}$ given by:
\begin{align}
	(Y,Z)=\begin{cases}
		(X,X) & \mathrm{w.p.\ } (1-p)(1-q_1)(1-q_2)\\
		(X,\perp) & \mathrm{w.p.\ } (1-p)(1-q_1)q_2\\
		(\perp,X) & \mathrm{w.p.\ } (1-p)q_1(1-q_2)\\
		(\perp,\perp) & \mathrm{w.p.\ } p + (1-p)q_1q_2
	\end{cases}
	\label{eq:eg1}
\end{align}
%Let $A$ be the binary random variable denoting the outcome of the BEC($p$) channel. That is, $A=1$ exactly when the BEC($p$) decides to erase. Similarly, let $B$ and $C$ denote the outcomes of the BEC($q_1$) and BEC($q_2$) channels respectively.\\\\
%\textbf{Achievability:}\\
\subsection{Inner Bound}
We want to construct a distribution $p_{XYZUVW}=q_Xp_{UVW|X}p_{Y|UV} p_{Z|UW}$ such that the marginal $p_{XYZ}=q_Xq_{YZ|X}$.\\
Let $A\sim\mathrm{Bernoulli}(p),B\sim\mathrm{Bernoulli}(q_1)$ and $C\sim\mathrm{Bernoulli}(q_2)$, with $A,B,C,X$ independent.\\
Now let $p_{UVW|X}$ be as follows:
\begin{align*}
	U={}& \begin{cases}
		\perp & \mathrm{if\ } A=1\\
		\mathrm{Bernoulli}(1/2) & \mathrm{if\ } A=0,B=C=1\\
		X & \mathrm{otherwise} 
	\end{cases}\\
	V={}& \begin{cases}
		\perp & \mathrm{if\ } A=1\\
		B & \mathrm{otherwise}
	\end{cases}\\
	W={}& \begin{cases}
		\perp & \mathrm{if\ } A=1\\
		C & \mathrm{otherwise}
	\end{cases}
\end{align*}
Finally, Set $p_{Y|UV},p_{Z|UW}$ to
\begin{align*}
	Y={}&\begin{cases}
		U & \mathrm{if\ } V=0\\
		\perp & \mathrm{otherwise}
	\end{cases}\\
	Z={}&\begin{cases}
		U & \mathrm{if\ } W=0\\
		\perp & \mathrm{otherwise}
	\end{cases}
\end{align*}
The conditional independence requirements are obviously met. To check that the marginal $p_{XYZ}$ is correct:\\
Notice that $V=0$ iff $A=B=0$ and $W=0$ iff $A=C=0$. Also, if $A=0$ and at least one of $B,C$ is 0, then $U=X$. Therefore, we have
\begin{align}
	\begin{split}
		Y={}&\begin{cases}
			X & \mathrm{if\ } A=B=0\\
			\perp & \mathrm{otherwise}
		\end{cases}\\
		Z={}&\begin{cases}
			X & \mathrm{if\ } A=C=0\\
			\perp & \mathrm{otherwise}
		\end{cases}
	\end{split}
	\label{eq:eg2}
\end{align}
It is easy to check that (\ref{eq:eg1}) and (\ref{eq:eg2}) are equivalent.\\\\
Calculating the various mutual information values we get
\begin{align*}
	I(V;W|U)={}& 0\\
	I(U,V,W;X) ={}& (1-p)(1-q_1q_2)\\
	I(U;X,Y,Z) ={}& (1-p)(1-q_1q_2) + I(A;Y,Z)\\
	I(U,V;X,Y,Z) ={}& (1-p)(1-q_1q_2) + I(A;Y,Z)\\
	& + (1-p)h(q_1)\\
	I(U,W;X,Y,Z) ={}& (1-p)(1-q_1q_2) + I(A;Y,Z)\\
	& + (1-p)h(q_2)\\
	I(U,V,W;X,Y,Z) ={}& (1-p)(1-q_1q_2) + I(A;Y,Z)\\
	& + (1-p) \left[h(q_1)+h(q_2)\right]\\
	\mathrm{where}\quad	I(A;Y,Z) ={}& h(p)-q_1q_2 h\left(\frac{p}{p+(1-p)q_1q_2}\right)
\end{align*}
The resulting inner bound on the rate region is
\begin{align*}
	R\ge{}& (1-p)(1-q_1q_2)+\frac{1}{2}I(A;Y,Z)\\
	R+R_1\ge{}& (1-p)(1-q_1q_2)+I(A;Y,Z)\\
	&+(1-p)h(q_1)\\
	R+R_2\ge{}& (1-p)(1-q_1q_2)+I(A;Y,Z)\\
	&+(1-p)h(q_2)\\
	R+R_1+R_2\ge{}& (1-p)(1-q_1q_2)+I(A;Y,Z)\\
	&+(1-p)\left[h(q_1)+h(q_2)\right]\\
	R+\frac{1}{2}(R_1+R_2)\ge{}& (1-p)(1-q_1q_2) + I(A;Y,Z)\\
	&+\frac{1}{2}(1-p)\left[h(q_1)+h(q_2)\right]
\end{align*}
If we have unlimited shared randomness, the minimum communication rate needed with this scheme is $(1-p)(1-q_1q_2) + \frac{1}{2}I(A;YZ)$. Thus, $R_{\mathrm{opt}}\le (1-p)(1-q_1q_2)+\frac{1}{2}I(A;Y,Z)$.
\subsection{Lower Bound}
The converse given by Theorem \ref{th:conv} is
\begin{align*}
	R_{\mathrm{opt}}\ge{}& \min_{p_{U|XYZ}}\max\Big\{I(Y;Z|U)+I(U,Y,Z;X)\;,\\
	&\frac{1}{2}\left[I(Y;Z|U)+I(U,Y,Z;X)+I(U;X,Y,Z)\right]\Big\}
\end{align*}
This optimisation problem seems difficult. The best lower bound we currently have is $R_{\mathrm{opt}}\ge I(X;Y,Z)=(1-p)(1-q_1q_2)$.
\section{Equivalent form of lower bound statement}\label{sec:app_equiv}
In this section, we will show that the lower bounds in (\ref{eq:conv5}) and Theorem \ref{th:conv} are equivalent\footnote{See also \cite[proof of lemma 1]{gowtham}}, that is,
\begin{align}
	&\min_{p_{U|XYZ}}\max\left\{I(Y;Z|U)+I(U,Y,Z;X),\right.\nonumber\\
	&\left.I(U;X,Y,Z)\right\}\nonumber\\
	&\min_{p_{U|XYZ}}\max\left\{I(Y;Z|U)+I(U,Y,Z;X),\right.\nonumber\\
	&\left.\frac{1}{2}\left(I(Y;Z|U)+I(UY,Z;X)+I(U;X,Y,Z)\right)\right\}
\end{align}
First, note that
\begin{align*}
	I(U,Y,Z;X)={}&I(U;X|Y,Z)+I(X;Y,Z)\\
	I(U;X,Y,Z)={}&I(U;X|Y,Z)+I(U;Y,Z)
\end{align*}
Thus, we want to show
\begin{align}
	&\min_{p_{U|X,Y,Z}}\left[I(U;X|Y,Z)+\right.\nonumber\\
	&\left.\max\left\{I(Y;Z|U)+I(X;Y,Z)\,,\,I(U;Y,Z)\right\}\right]\nonumber\\
	={}& \min_{p_{U|X,Y,Z}}\left[I(U;X|Y,Z)+\max\left\{I(Y;Z|U)+I(X;Y,Z),\right.\right.\nonumber\\
	&\left.\left.\frac{1}{2}\left(I(Y;Z|U)+I(X;Y,Z)+I(U;Y,Z)\right)\right\}\right]\label{eq:app:d}
\end{align}
We will use the shorthand $f_1(p):=I(Y;Z|U)+I(X;Y,Z), f_2(p):=I(U;Y,Z),f_3(p):=I(U;X|Y,Z)$, where $p=p_{U|XYZ}$ is the distribution of $U$.

In showing (\ref{eq:app:d}), L.H.S$\,\ge\,$R.H.S. is clear. To show L.H.S$\,\le\,$R.H.S., it is enough to show that there exists a minimiser $p_{min}$ of R.H.S of (\ref{eq:app:d}) such that $f_1(p_{min})\ge 0.5(f_1(p_{min})+f_2(p_{min}))$.

We start with a minimiser $U_0\sim p_{U|XYZ}^*=:p^*$ of R.H.S. of (\ref{eq:app:d}). Let $U_1$ be a constant random variable and let $Q\sim\mathrm{Bernoulli}(\theta)$, where $\theta\in[0,1]$. Now, let the distribution $U'=(U_Q,Q)$ be denoted by $p_{U|XYZ}{}'=:p'$. We will show that $p'$ for some value of $\theta$ is the required minimiser.

Note that:
\begin{align*}
	f_1(p')={}&I(Y;Z|U')+I(X;Y,Z)\\
	={}&I(Y;Z|U_Q,Q)+I(X;Y,Z)\\
	={}& (1-\theta)f_1(p^*)+\theta (I(Y;Z)+I(X;Y,Z))\\
	f_2(p')={}&I(U';Y,Z)\\
	={}& (1-\theta) I(U_0;Y,Z)\\
	={}&(1-\theta)f_2(p^*)\\
	f_3(p')={}&I(U';X|Y,Z)\\
	={}& (1-\theta )I(U_0;X|Y,Z)\\
	={}&(1-\theta)f_3(p^*)
\end{align*}

Assume w.l.o.g. that $p^*$ is \emph{not} the required minimiser. That is, $f_1(p^*)<0.5(f_1(p^*)+f_2(p^*))$.

Note that since $I(Y;Z|U)+I(U;Y,Z)=I(Y;Z)+I(Y;U|Z)+I(Z;U|Y)$,
\begin{align*}
	&f_1(p')+f_2(p')\\
	={}&I(X;Y,Z)+I(Y;Z)+I(Y;U_Q,Q|Z)+I(Z;U_Q,Q|Y)\\
	={}&I(X;Y,Z)+I(Y;Z)+I(Y;U_Q|Z,Q)+I(Z;U_Q|Y,Q)\\
	={}& I(X;Y,Z)+I(Y;Z)+(1-\theta)(I(Y;U_0|Z)+I(Z;U_0|Y))\\
	&\le{}f_1(p^*)+f_2(p^*)
\end{align*}
If $I(Y;U_0|Z)+I(Z;U_0|Y)>0$, then we have a strict inequality above if $\theta>0$. Also, we can get $f_1(p')\le f_1(p^*)+\delta$ for any $\delta>0$ by choosing $\theta>0$ sufficiently small.
Thus, if $f_1(p^*)<0.5(f_1(p^*)+f_2(p^*))$, then $f_1(p')<0.5(f_1(p^*)+f_2(p^*))$. Thus, we have $f_1(p')<f_1(p^*)$, $f_1(p')+f_2(p')<f_1(p^*)+f_2(p^*)$, and $f_3(p')<f_3(p^*)$. This contradicts the fact that $p^*$ is a minimiser.

Thus, we must have $I(Y;U_0|Z)+I(Z;U_0|Y)=0$. Then, $f_1(p')+f_2(p')=f_1(p^*)+f_2(p^*)=I(Y;Z)+I(X;Y,Z)$.
This implies $f_1(p')=(1-\theta)f_1(p^*)+\theta(f_1(p^*)+f_2(p^*))$. Since $f_1(p^*)<0.5(f_1(p^*)+f_2(p^*))<f_1(p^*)+f_2(p^*)$, by the intermediate value theorem, there is a $\theta$ such that $f_1(p')=0.5(f_1(p^*)+f_2(p^*))=0.5(f_1(p')+f_2(p'))$. This is the required minimiser, since $f_1(p')=\max\{f_1(p'),0.5(f_1(p')+f_2(p'))\}=0.5(f_1(p^*)+f_2(p^*))>f_1(p^*)$.

\end{document}